%% file: author.tex
\documentclass[runningheads]{svmult}

\usepackage{makeidx}   
\usepackage{graphicx}  
\usepackage{subeqnar}  
\usepackage{multicol}  
\usepackage{physmubb}  
\makeindex             

\include{DEFS}
\begin{document}
\title*{Measurements of diffractive processes at HERA}
%
%
%
%
\titlerunning{Diffractive processes at HERA}
%
\author{Aharon Levy\\ for the H1 and ZEUS Collaborations
}
\authorrunning{Aharon Levy}
%
%

\maketitle              

\section{Introduction}
One of the main objectives of $e p$ scattering is to study the
structure of the proton. This is done by performing a deep inelastic
scattering (DIS) process in which the virtual photon probes the partonic
structure of the proton. All events are studied, like in a total cross
section experiment.

If, however, the final state consists of events in which the proton
remains intact, or there is a large rapidity gap (LRG), we have
diffractive events, mediated by a color singlet exchange. These can be
further classified as inclusive diffractive events, as depicted in
Fig. 1, or as exclusive ones, shown in Fig. 2. In this talk, these two
classes of processes will be studied.

\begin{figure}
\begin{minipage}{5.6cm}
\begin{center}
\includegraphics[width=.6\textwidth]{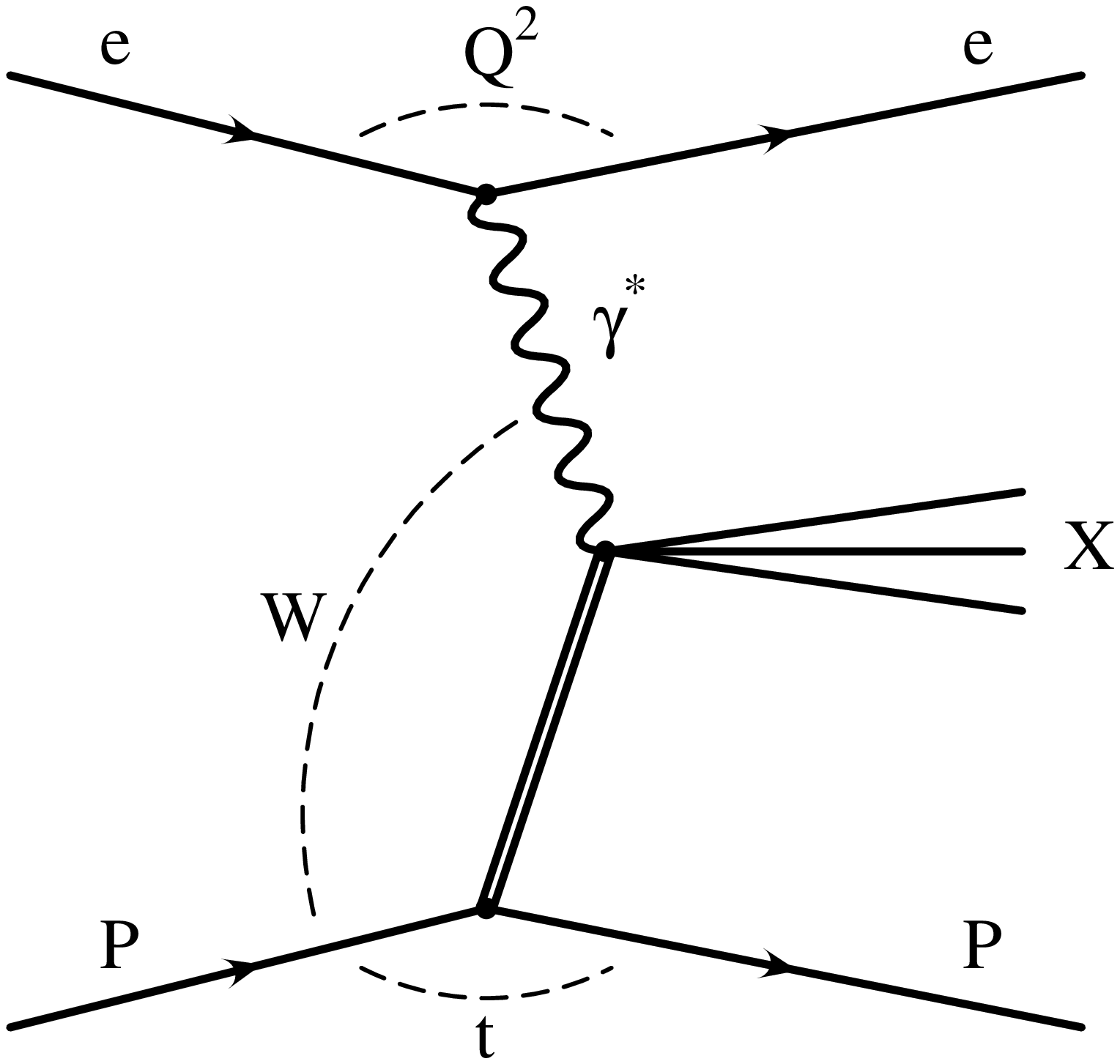}
\caption{Schematic diagram for inclusive diffractive DIS $ep$
reaction, where the LRG is between the proton and the final state $X$.}
\label{fig:diagincl}
\end{center}
\end{minipage}
\hspace*{3mm}
\begin{minipage}{5.6cm}
\begin{center}
\includegraphics[width=.6\textwidth]{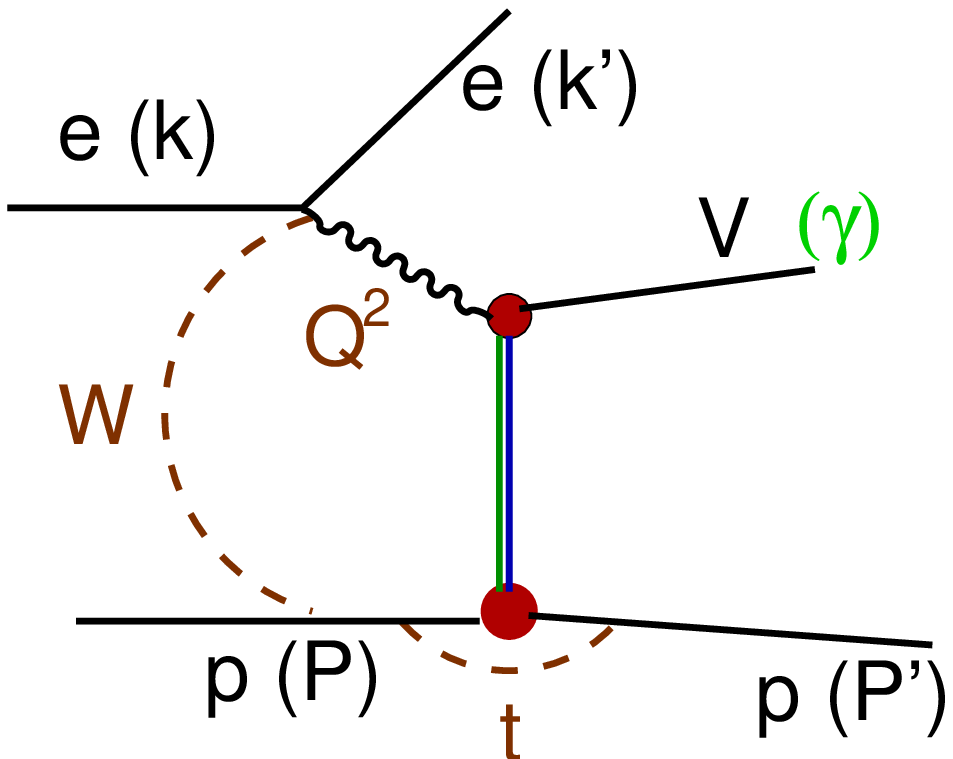}
\caption{Schematic diagram for exclusive diffractive electroproduction
of vector mesons $V$, where the rapidity gap is between $V$ and $p$.} 
\label{fig:diagexcl}
\end{center}
\end{minipage}
\end{figure}

In the Regge theory of strong interactions~\cite{regge}, the color
singlet exchange is due to the soft Pomeron, $\pom$, introduced by
Gribov~\cite{gribov}, and the parameters of its trajectory have been
determined by Donnachie and Landshoff~\cite{dl}. In contrast to the
universal nature of this exchange, in the language of Quantum
Chromodynamics (QCD), the color singlet exchange is described as a two
gluon exchange~\cite{low-nussinov}. Since its properties depend on the
scale involved in the interaction, the QCD Pomeron has a non-universal
character. We will discuss the results with respect to the soft and
the hard behavior of these two approaches, which manifest themselves
in the energy behavior of the cross sections.

\section{Kinematics of diffractive scattering}
The variables used in diffractive scattering can be defined using the
four vectors presented in Fig. 3. The usual DIS variables are the
negative of the mass squared of the virtual photon, $Q^2 = -q^2 = -(k
-k')^2$, the square of the center of mass energy of the $\gamma*p$
system,  $W^2 = (q + p)^2$, the Bjorken scaling variable, 
$x =\frac{\textstyle Q^2}{\textstyle 2 p \cdot q}$, which in the Quark
Parton Model can be thought of the fraction of the proton momentum
carried  by the interacting quark, and the inelasticity, 
$y =\frac{\textstyle q \cdot p}{\textstyle k \cdot p}$.

\vspace*{-0.5cm}
\begin{figure}
\begin{minipage}{5.5cm}
\begin{center}
\includegraphics[width=.55\textwidth]{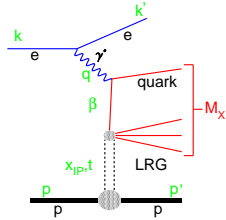}
\caption{Schematic diagram of diffractive $ep$ interaction. }
\label{fig:kin}
\end{center}
\end{minipage}
\hspace*{3mm}
\begin{minipage}{5.5cm}
In addition to the above variables, the variables used to describe
the diffractive final state are
\begin{eqnarray*}
\xpom &=& \frac{q\cdot (p-p')}{q\cdot p}\simeq
\frac{Q^2+M_X^2}{Q^2+W^2}   \\
\beta &=& \frac{Q^2}{2q\cdot (p-p')}\simeq \frac{Q^2}{Q^2+M_X^2}  \\
t &=& (p-p')^2.
\end{eqnarray*}
\end{minipage}
\end{figure}

\vspace*{-0.5cm}
\noindent The fractional proton momentum which participates in the interaction
with $\gamma*$ is $\xpom$, and $\beta$ is the equivalent of Bjorken
$x$ but relative to the exchanged object. $M_X$ is the invariant mass
of the hadronic final state recoiling against the leading proton,
$M_X^2 = (q + p - p^\prime)^2$. The approximate relations hold for
small $t$ and large $W$.

\section{Diffraction as soft or hard process}

In the Regge description, diffraction is a soft process. Its properties
are determined by the universal Pomeron trajectory, $\apom(t) =
\apom(0) + \aprime t$, with $\apom(0)$ = 1.081 and $\aprime$ = 0.25
GeV$^{-2}$. Therefore the energy behavior of the total
$\gamma^*p$ cross section is expected to be $\sigma_{tot} \sim
(W^2)^{\apom(0)-1} \simeq W^{0.16}$. Both the elastic and the
inclusive diffraction cross section are expected to have a faster rise
with energy $\sim (W^2)^{2\apom(0)-2}/b$, where $b$ is the slope of
the differential cross section in $t$~\cite{ha-ferrara}.

In the perturbative QCD picture, the diffractive process is viewed, in
the proton rest frame, as follows: the virtual photon fluctuates into
a quark-antiquark pair, which interact diffractively with the proton by
exchanging two gluons. Therefore, in this case, the diffractive cross
section is proportional to the square of the gluon density. Since the
gluon density, $xg(x,Q^2) \sim x^{-\lambda} \sim (W^2)^\lambda$, where
$\lambda$ depends on the scale $Q^2$, the diffractive cross section
has an energy behavior $\sim (W^4)^{\lambda(Q^2)}$. At $Q^2$ = 10
GeV$^2$, for example, $\lambda \approx$ 0.2, and thus the cross
section would have a $W^{0.8}$ dependence.

The above discussion shows that we expect a transition from soft to hard
processes when the virtuality of the probing photon increases.

\section{Inclusive diffraction}

One can express the inclusive cross section by a diffractive structure
function $F_2^D$ which is a function of four variables, $\xpom, t, x,
Q^2$. It was shown~\cite{collins,berera,trentadue} that QCD
factorization holds also in case of diffraction. Thus $F_2^D$ can be
decomposed into diffractive parton distributions, which would follow
the same DGLAP evolution equation that apply in the DIS inclusive
case. If, in addition, one postulates Regge factorization, in the
spirit of Ingelman and Schlein~\cite{ingelman}, $F_2^D$ may be
decomposed into a universal $\pom$ flux and the structure function of
the $\pom$. One usually integrates over the $t$ variable, and
this decomposition is written as
\[
\frac{dF_2^{D(3)}(x,Q^2,\xpom)}{d\xpom} =
f\pomsub(\xpom)F_2^{\pom}(\beta,Q^2),
\] 
where the $\xpom$ dependence of the flux is universal, independent of
$\beta$ and $Q^2$ and is given by $f\pomsub(\xpom) \sim
\xpom^{1-2\apom(0)}$. 
\vspace{-1cm}
\begin{figure}[h]
\begin{minipage}{5.5cm}
\includegraphics[width=1.2\textwidth]{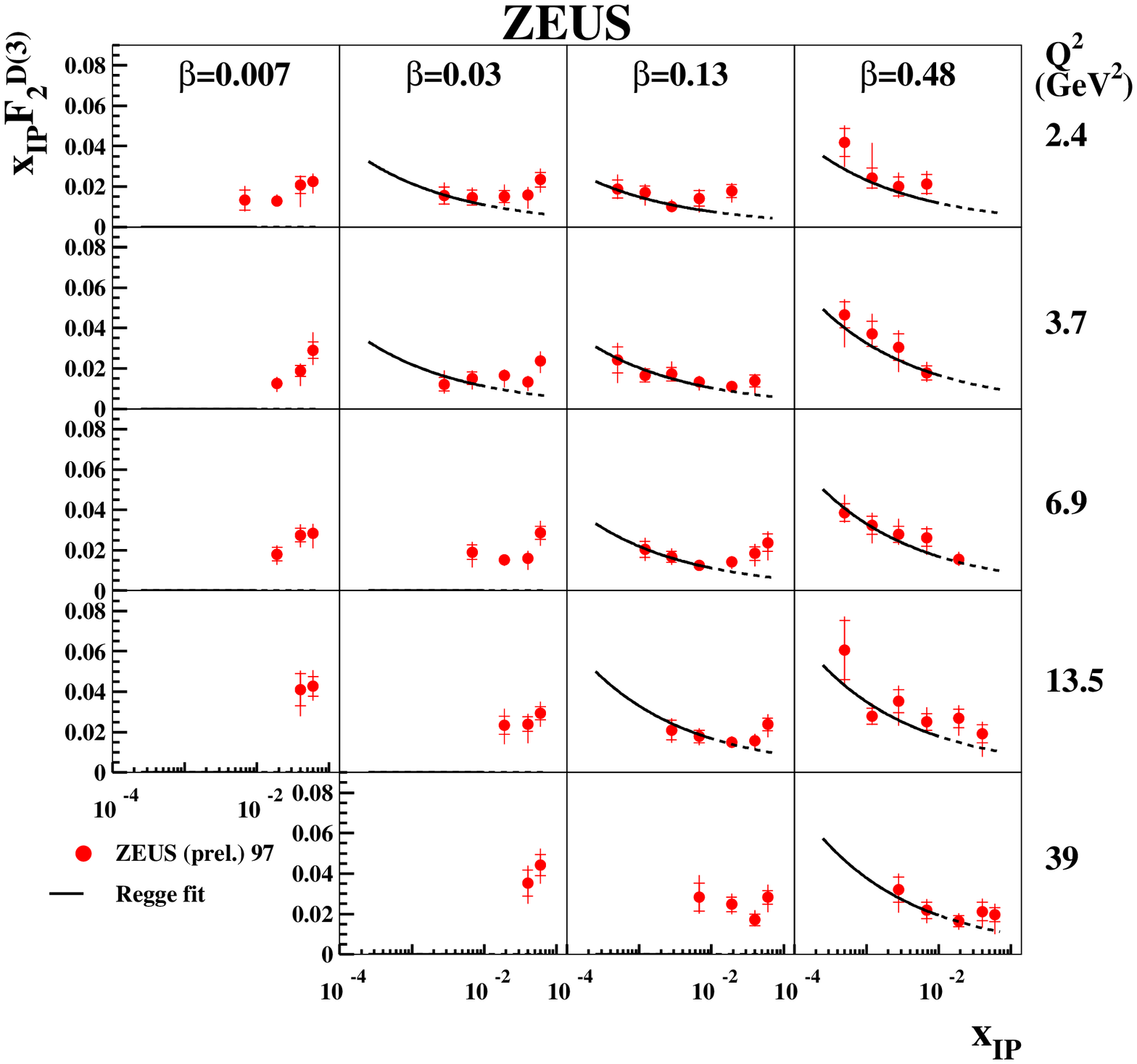}
\caption{$\xpom$ dependence of $\xpom F_2^{D(3)}$ at fixed $\beta$ and
$Q^2$ values, as denoted in the figure.}
\label{fig:f2d3}
\end{minipage} 
\hspace{7mm}
\begin{minipage}{5.5cm}
\vspace{-2cm}
\includegraphics[width=1.5\textwidth]{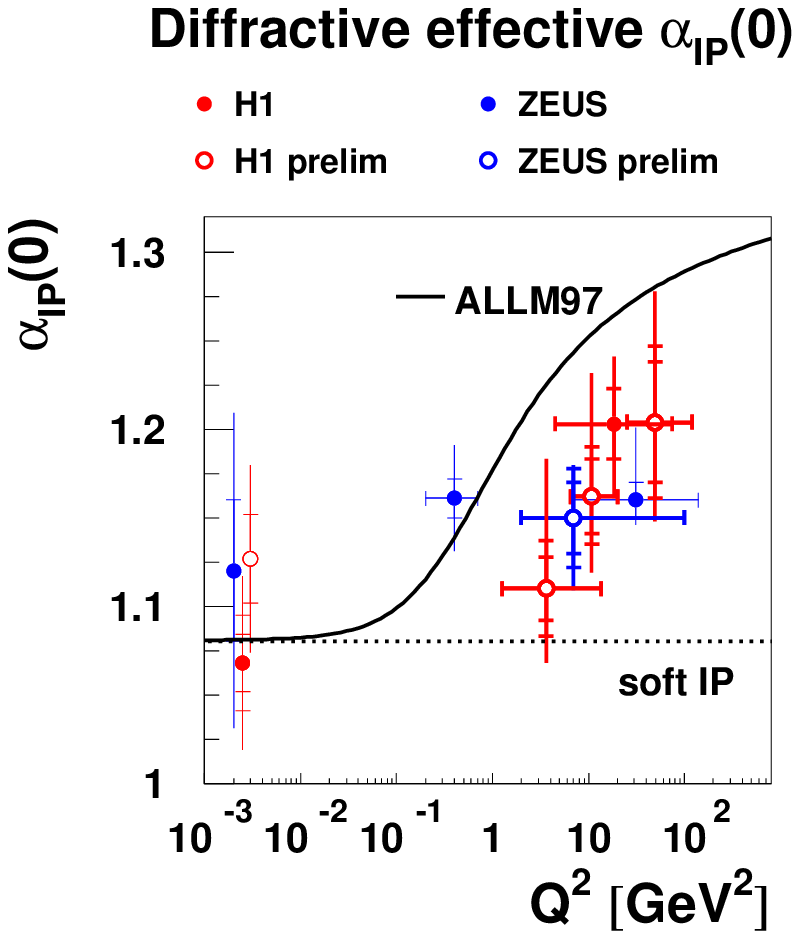}
\caption{$Q^2$ dependence of $\apom(0)$ derived from measurements of
diffractive and total $\gamma^*p$ cross sections. The curve (ALLM97)
is a representation of the results obtained in inclusive DIS measurements.}
\label{fig:apom}
\end{minipage}
\end{figure}

Fig. 4 shows the $\xpom$ dependence of $\xpom F_2^{D(3)}$ at fixed
$\beta$ and $Q^2$ values, as measured by the ZEUS
collaboration~\cite{zeus-f2d3}. The curves are the best fit to the
data (restricted to $\xpom < 0.01$ using a universal flux, as
described above, with $\apom$ = 1.16 $\pm$
.01(stat)$^{+.04}_{-.01}$(syst). This value, together with a
compilation from other measurements~\cite{h1-pom}, is displayed in
Fig. 5. For comparison, also shown is the $Q^2$ dependence of
$\apom(0)$ derived from the inclusive DIS measurements and
conveniently represented by the ALLM97 parameterization~\cite{allm97}. 
Clearly, the inclusive DIS data are not compatible with a universal
$\pom$ trajectory. The diffractive measurements seem to point to some
$Q^2$ dependence, though the uncertainties are too large for a firm
conclusion. For $Q^2 >$ 10 GeV$^2$, the value of $\apom$(0) is
significantly higher than that expected from the soft Pomeron.

The $\beta$ and $Q^2$ dependence of the Pomeron structure function, as
measured by the H1 collaboration~\cite{h1-pom},
are shown in Fig. 6. It is the Pomeron structure function under the
assumption that the longitudinal diffractive structure function is
zero, $F_L^D$ = 0, and that Regge factorization holds, and therefore
one divides-out the Pomeron flux. One sees that, just like in the
inclusive DIS case, as $Q^2$ increases, the Pomeron structure function
is consistent with a rising behavior towards low $\beta$. However,
unlike the inclusive DIS case, positive scaling violations are
observed up to large $\beta$ values, and only for $\beta >$ 0.6, the
scaling violations turn negative.
\vspace{-0.5cm}
\begin{figure}[h]
\hspace{-1cm}
\includegraphics[width=.6\textwidth]{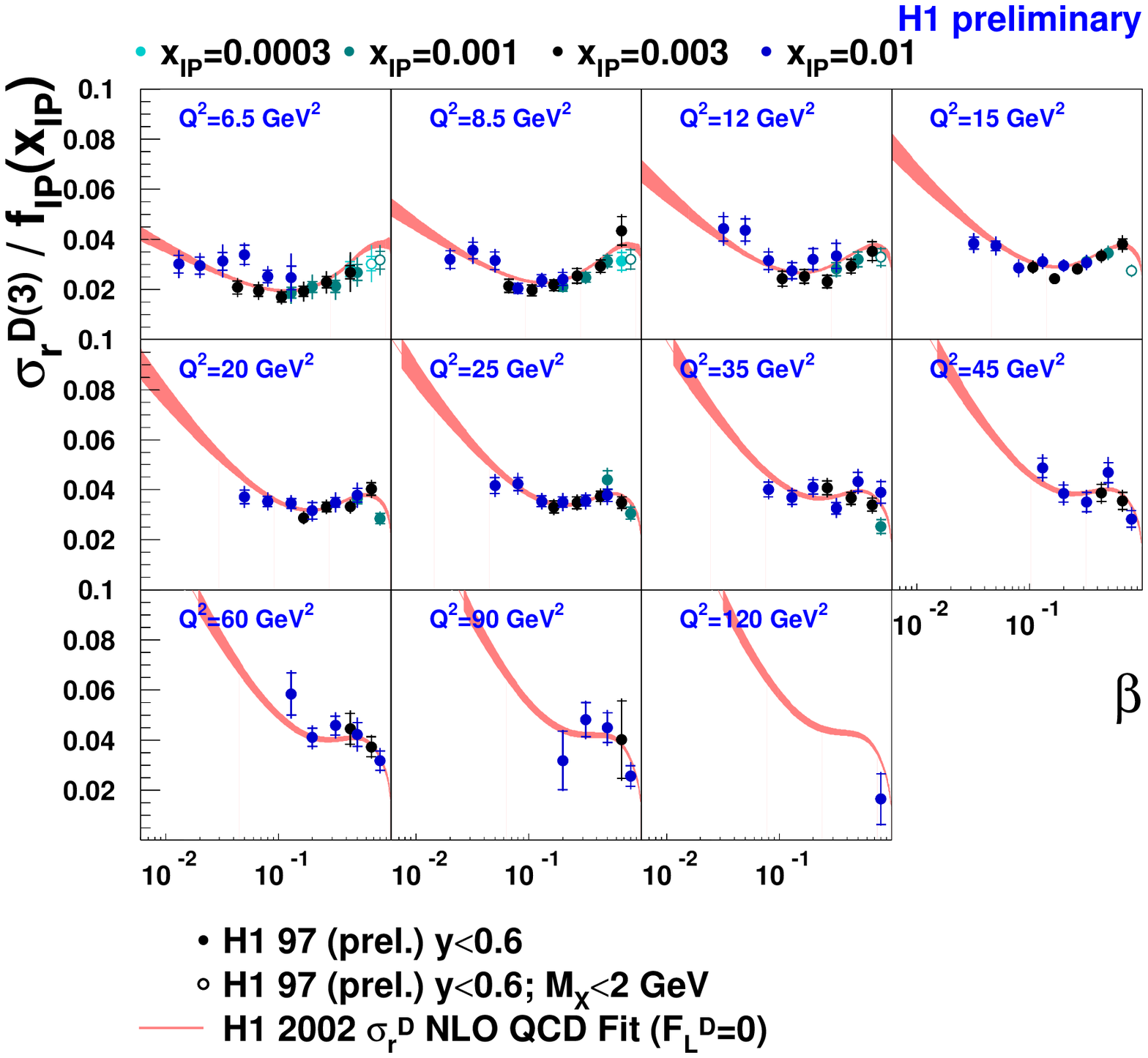}
\includegraphics[width=.6\textwidth]{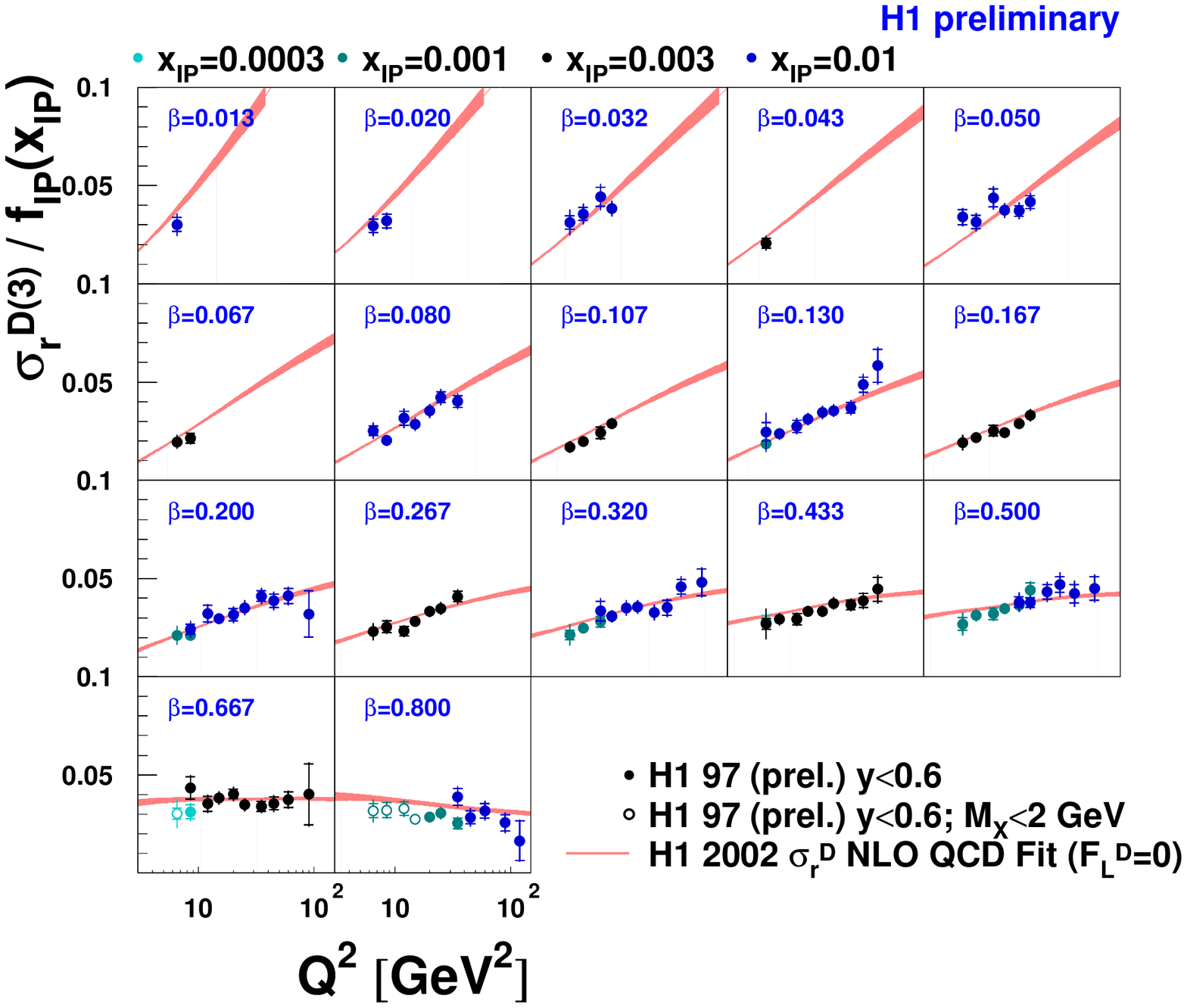}
\vspace{-1.5cm}
\caption{The Pomeron structure function dependence on $\beta$ (left)
and on $Q^2$ (right). The curves are a result of a
NLO QCD fit, assuming $F_L^D$ = 0.}
\label{fig:pom}
\end{figure}

\vspace{-0.4cm}
An NLO QCD fit, assuming $F_L^D$ = 0, was performed to the data and a
good description of the data is obtained. The resulting parton density
distributions in the Pomeron are shown in Fig. 7. They do not differ
much from a LO QCD fit, and have the feature of a sizable contribution
of the gluon density at large $z$ (which is same as $\beta$). Using
the parton densities from the NLO fit, one gets a good description of
the $\beta$ and $\xpom$ distribution of diffractive jet
production~\cite{h1-jet} and diffractive $D^*$
production~\cite{zeus-dstar}. 

\begin{figure}
\begin{minipage}{5.5cm}
\includegraphics[width=1.2\textwidth]{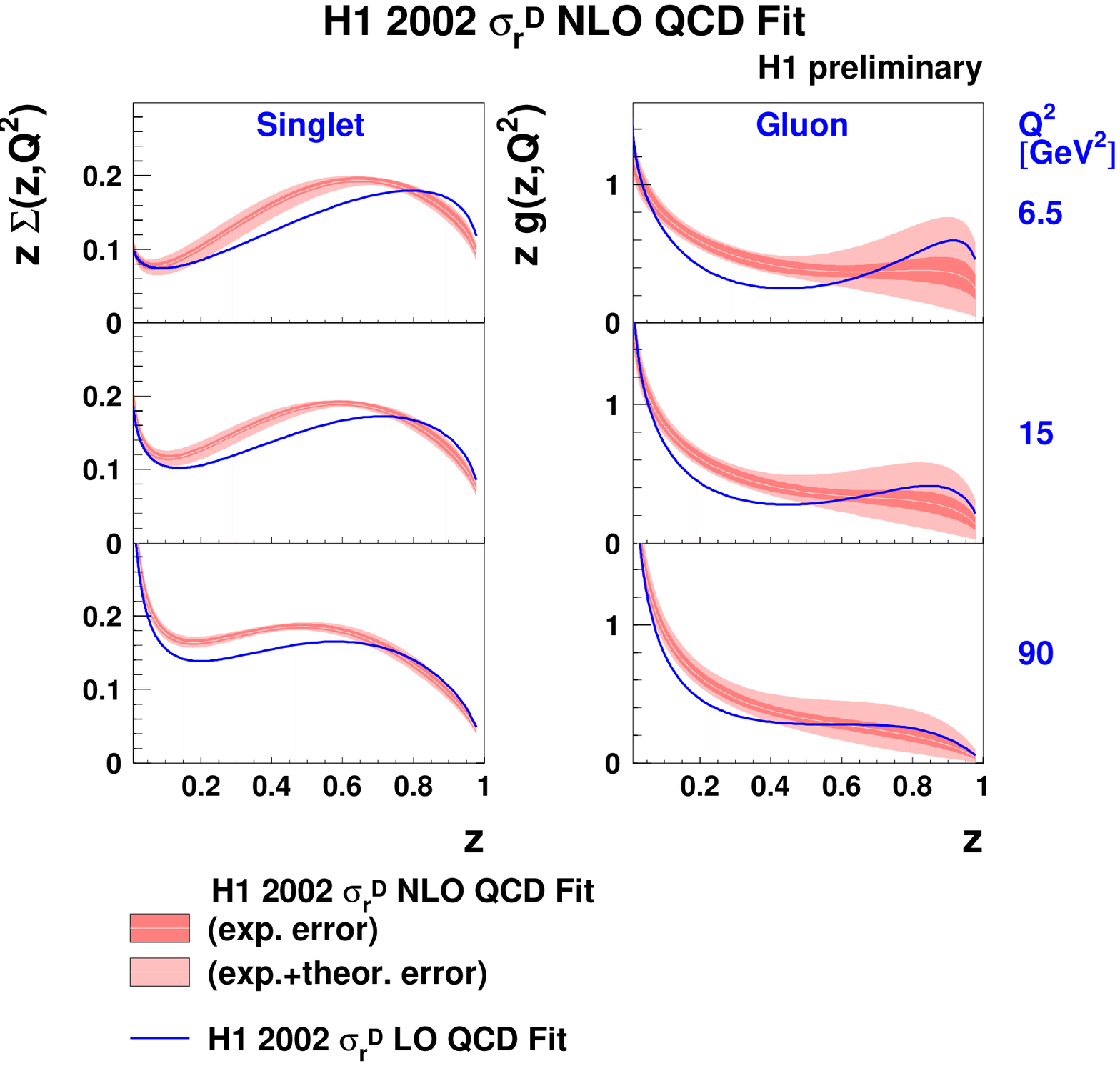}
\caption{The resulting parton density distributions in the Pomeron, using a
NLO QCD fit (shaded line) compared to a LO fit (solid line).}
\end{minipage}
\hspace{5mm}
\begin{minipage}{5.5cm}
\vspace{-0.4cm}
\includegraphics[width=1.2\textwidth]{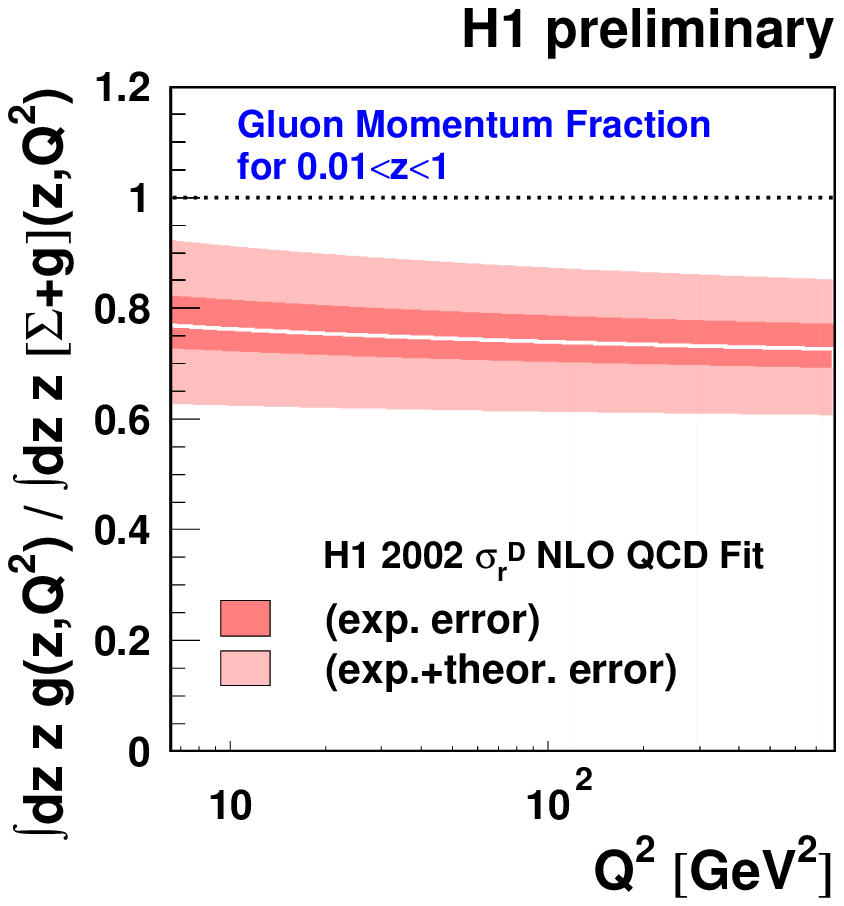}
\caption{The gluon momentum fraction from a NLO QCD fit, as function
of $Q^2$.}
\label{fig:gluon-frac}
\end{minipage}
\end{figure}
One can calculate the momentum fraction taken by the gluons. This
turns out to be a large fraction, about 0.75 $\pm$ 0.15 at $Q^2$ = 10
GeV$^2$, and almost $Q^2$ independent, as can be seen in
Fig. 8. Frankfurt and Strikman~\cite{fs} used this to calculate the
probability that a gluon from the proton will produce a diffractive
process. The find that at $x = 10^{-3}$ and $Q^2$ = 4 GeV$^2$, this
probability is as high as 0.4, which is very close to the unitarity
limit of 0.5.

The ratio of the diffractive cross section to the total $\gamma*p$
cross section is shown in Fig. 9, as function of $Q^2$, for fixed
$\beta$ values. This ratio, is remarkably flat over a wide kinematic range.
The ratio is flat also as function of $W$ for fixed $M_X$
values~\cite{zeus-mx}, contrary to expectations from Regge phenomenology.
\vspace{-0.8cm}
\begin{figure}[h]
\begin{center}
\includegraphics[width=.65\textwidth]{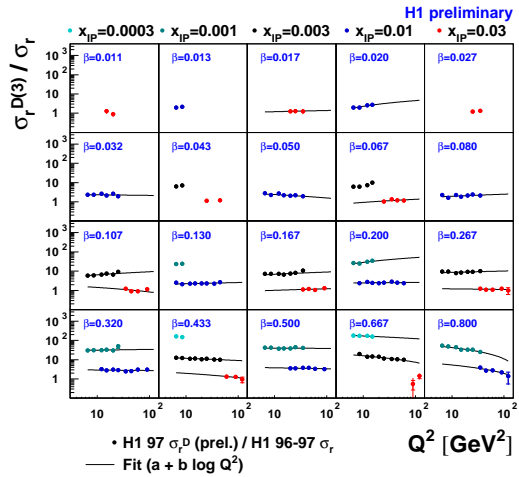}
\end{center}
\vspace{-1.2cm}
\caption{The ratio of the diffractive to total cross section as
function of $Q^2$, for fixed $\beta$ values.}
\label{fig:ratio-incl}
\end{figure}

\section{Exclusive Vector Mesons}

It has been suggested~\cite{brodsky} that a good way to see more
clearly the different behavior of soft and hard processes is to study
diffractive production of low masses, in particular vector mesons.
Exclusive vector meson (VM) production shows a clear interplay between
soft and hard diffractive processes. One of the nice examples to this
effect can be seen in case of the elastic photoproduction of VMs,
whose cross section measurements as function of $W$ are presented in
Fig. 10.  There is a clear change in the $W$ dependence when going
from the light VMs, like $\rho^0, \omega$ and $\phi$, to the heavier
ones like $J/\psi, \psi$(2S) and $\Upsilon$. In the latter case, a
hard scale is provided by the mass of the heavy quarks. The shallower
$W$ dependence of the light VMs is consistent with that expected from
a soft process, mediated by the $\pom$ trajectory, while the steep $W$
dependence in case of the heavy VMs is consistent with expectations
from a two-gluon exchange hard diffractive process calculated in pQCD.
\vspace{-1cm}
\begin{figure}
\begin{center}
\includegraphics[width=.8\textwidth]{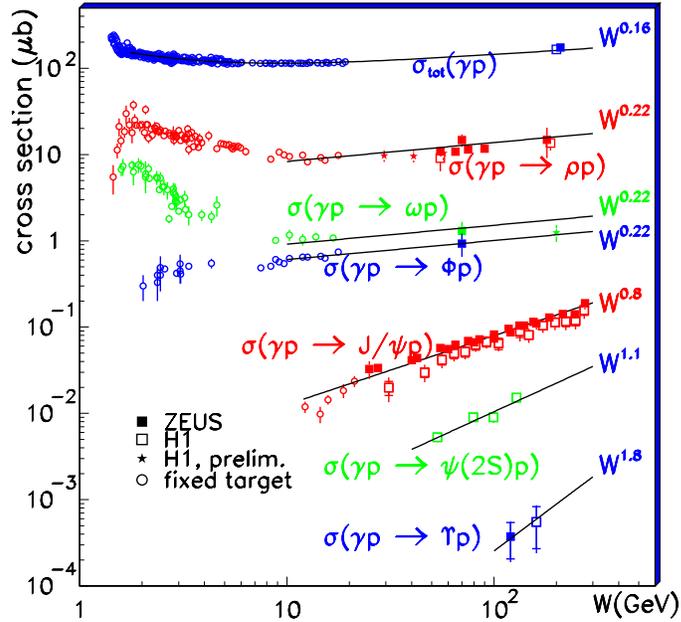}
\end{center}
\vspace{-0.5cm}
\caption{Compilation of elastic photoproduction of vector mesons, as
function of $W$. The total $\gamma p$ cross section is plotted for
comparison.}
\label{fig:xsect}
\end{figure}

Another soft-hard transition can be obtained by using $Q^2$ as a scale
in case of exclusive electroproduction of
$\rho^0$~\cite{zeus-rho,h1-rho}. This is demonstrated in Fig. 11,
where the $\delta$ parameter, of the $W^\delta$ behavior of the cross
section, is plotted as function of $Q^2$. While at low $Q^2$, $\delta$
is consistent with expectations of a soft process, at higher $Q^2$ the
values of $\delta$ reach those expected from a hard process.

\begin{figure}
\begin{minipage}{5.5cm}
\hspace{-1.2cm}
\includegraphics[width=1.2\textwidth]{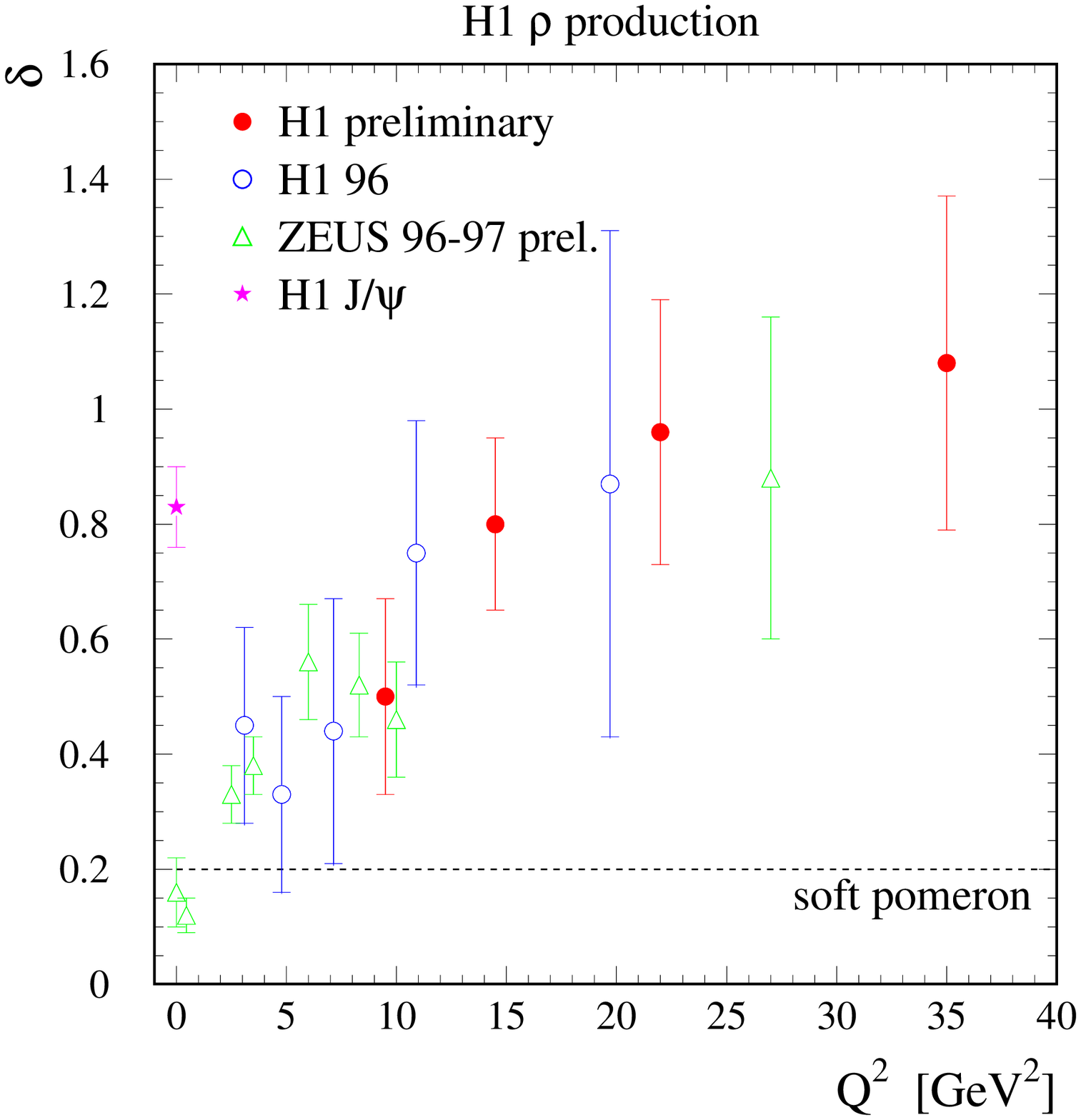}
\caption{The parameter $\delta$ from a fit of the form $W^\delta$ to the 
cross section data of $\rho^0$ electroproduction, as function of $Q^2$.}
\label{fig:delrhp}
\end{minipage}
\hspace{3mm}
\begin{minipage}{5.5cm}
\includegraphics[width=1.2\textwidth]{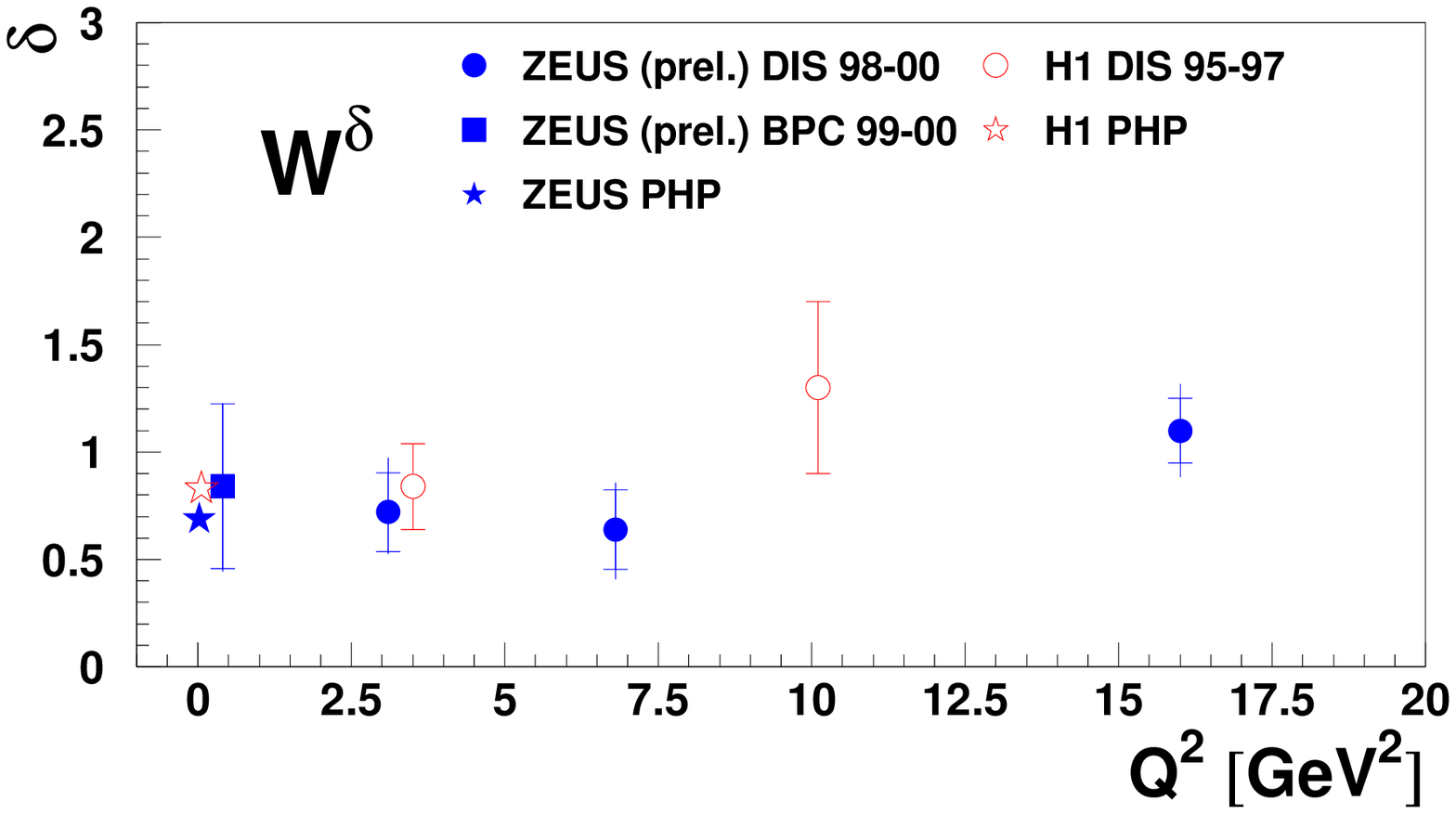}
\caption{The parameter $\delta$ from a fit of the form $W^\delta$ to 
the cross section data of $J/\psi$ electroproduction, as function of $Q^2$.}
\label{fig:delpsi}
\end{minipage}
\end{figure}

However, in case of exclusive electroproduction of
$J/\psi$~\cite{zeus-psi,h1-psi}, the hard scale is provided by the
heavy quark mass and thus the value of $\delta$ is already large even
at $Q^2$ = 0, as shown in Fig. 12.

\vspace{-0.5cm}
\begin{figure}
\begin{minipage}{5.5cm}
\includegraphics[width=1.\textwidth]{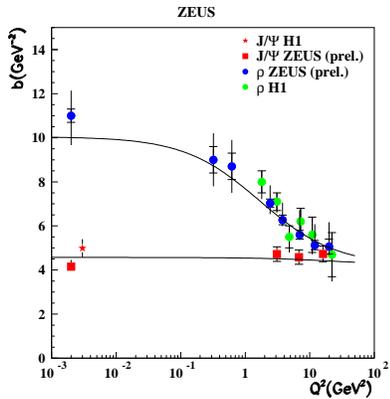}
\caption{The slope $b$ of $d\sigma/dt$ for $\rho^0$ and $J/\psi$.}
\label{fig:b}
\end{minipage}
\hspace{5mm}
\begin{minipage}{5.5cm}
Another way of seeing this different behavior of the light and heavy
vector mesons, is through the study of the $Q^2$ dependence of the
slope $b$ of the differential cross section $d\sigma/dt$ of $\rho^0$
and $J/\psi$. Fig. 13 displays the measured value of $b$ as function
of $Q^2$, for both vector mesons. One sees the clear
soft-hard transition in case of the $\rho^0$, while the $J/\psi$
production is a hard process even in case of photoproduction. At $Q^2
\geq$ 20 GeV$^2$, both mesons have the same small size, and the $b$
value is as expected from the proton size~\cite{afs}. 
\end{minipage}
\end{figure}

\vspace{-0.5cm}
Contrary to the photoproduction case, in the electroproduction of vector
mesons both transversely and linearly polarized photons participate. In
the picture discussed above, where the photon fluctuates into a
quark-antiqark dipole, it can do so in two configurations: a large spatial
one, resulting in a soft process, and a small spatial one, resulting in a
hard process~\cite{align-jet}. While the longitudinal photon is believed
to fluctuate into a small configuration, the transverse photon can
fluctuate into both. It is of interest to study how the different
configurations of the virtual photon influence the soft-hard transition
discussed above. To this end, one can use s-channel helicity conservation
to measure the ratio $R = \sigma_L/\sigma_T$ of the cross sections
produced by longitudinal to transverse photons.

\vspace{-0.5cm}
\begin{figure}
\begin{minipage}{5.5cm}
\hspace{-1.2cm}
\includegraphics[width=1.2\textwidth]{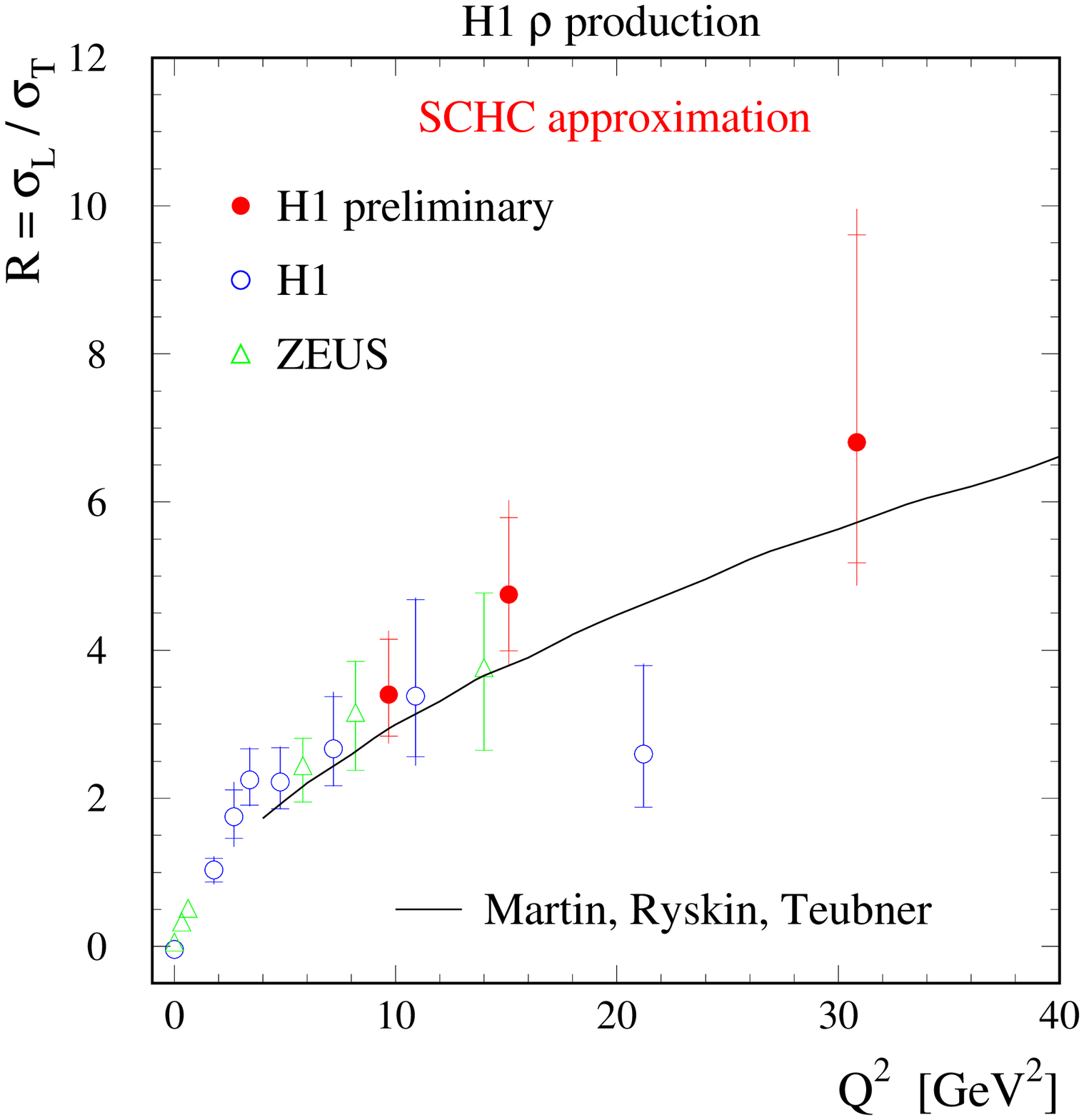}
\caption{The ratio $R$ as function of $Q^2$ for $\rho^0$
electroproduction. The curve is the expectation of the MRT
model~\cite{mrt}.}
\label{fig:rq2}
\end{minipage}
\hspace{3mm}
\begin{minipage}{5.5cm}
\hspace{-1cm}
\includegraphics[width=1.2\textwidth]{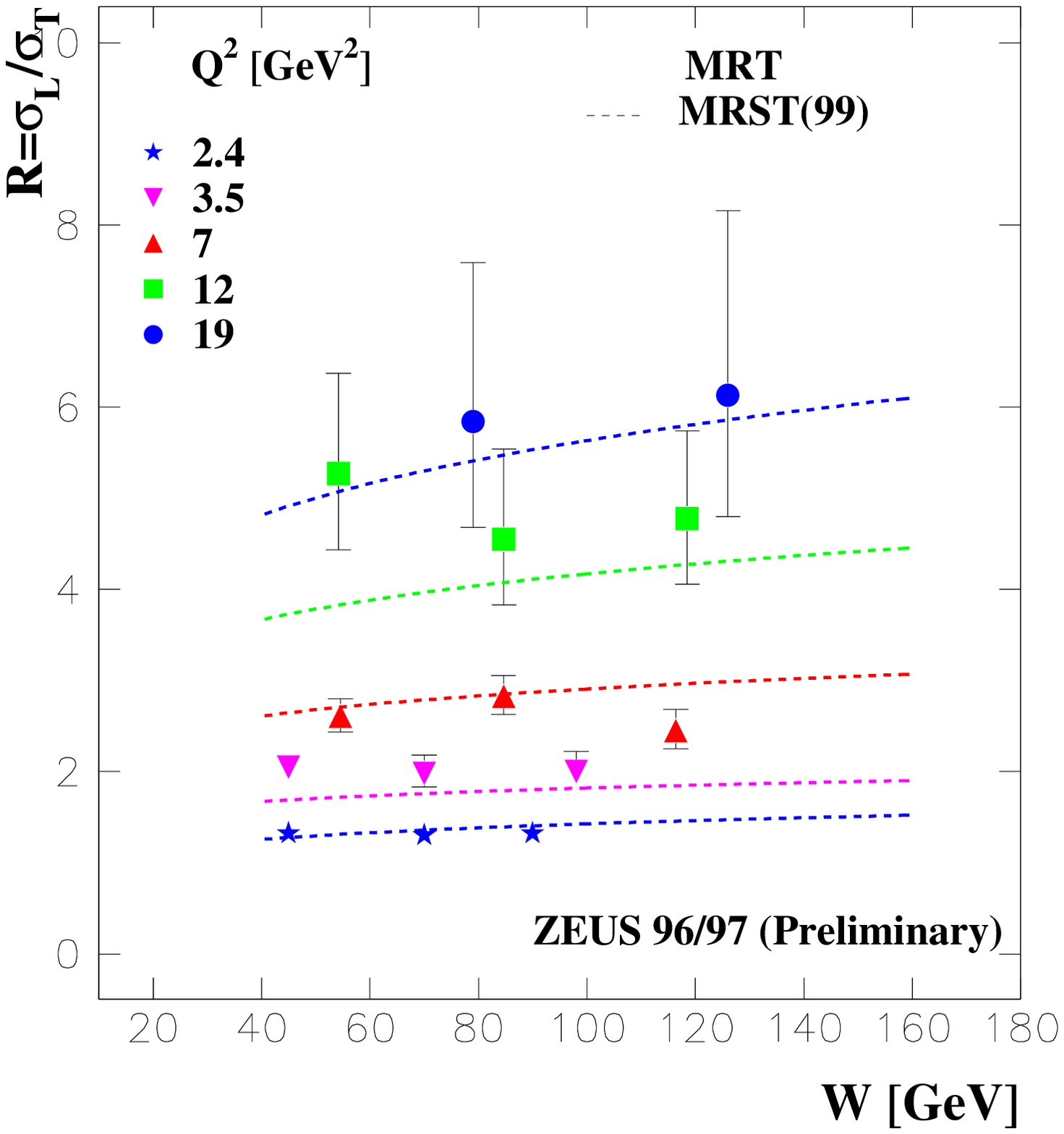}
\caption{The $W$ dependence of $R$ for $\rho^0$ electroproduction, 
for fixed $Q^2$ bins. The curves are the predicions of the MRT
model~\cite{mrt}.}
\label{fig:rw}
\end{minipage}
\end{figure}

Fig. 14 shows the ratio $R$ for electroproduction of $\rho^0$, as
function of $Q^2$~\cite{zeus-rho,h1-rq2}. The cross section coming
from the longitudinal photon dominates as $Q^2$ gets larger, and this
increase is well described by the MRT model~\cite{mrt}. What is
surprising is the fact that $R$ seems to be independent of $W$, in the
$Q^2$ range where the measurements were performed, as shown in
Fig. 15~\cite{zeus-rho}. This means that the $W$ dependence of
$\sigma_T$ is the same as $\sigma_L$, from which one concludes that
the large size configurations of the transverse photon are suppressed 
for $\rho^0$ electroproduction. This behavior is well reproduced in
the MRT model, even for the low $Q^2$ data.

Another striking result in case of the electroproduction of $\rho^0$
is shown in Fig. 16, where the ratio of the electroproduction to the
total $\gamma*p$ cross sections is displayed. This ratio is $W$
independent over the whole measured kinematic
region~\cite{zeus-ratio}. This is contrary to expectations of the
Regge approach as well as the pQCD one.
In case of the $J/\psi$, the ratio increases with $W$, and the
increase is consistent with expectations from both approaches.

\vspace{-0.5cm}
\begin{figure}
\begin{minipage}{5.5cm}
\hspace{-0.9cm}
\includegraphics[width=1.2\textwidth]{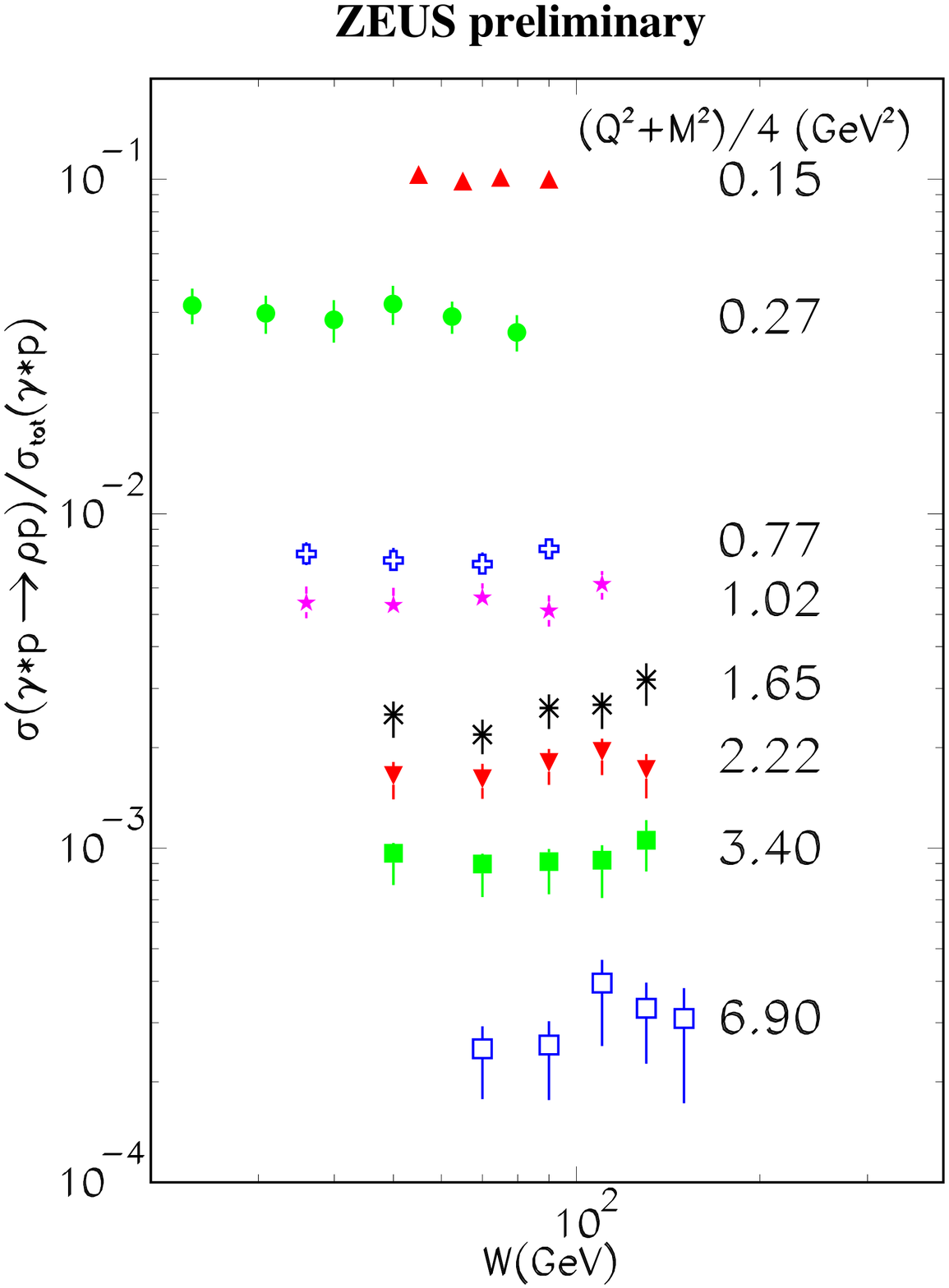}
\caption{The ratio of the $\rho^0$ electroproduction cross section to
the total $\gamma*p$ one, as function of $W$, at different scales.}
\label{fig:ratio-rho}
\end{minipage}
\hspace{5mm}
\begin{minipage}{5.5cm}
\hspace{-0.5cm}
\includegraphics[width=1.2\textwidth]{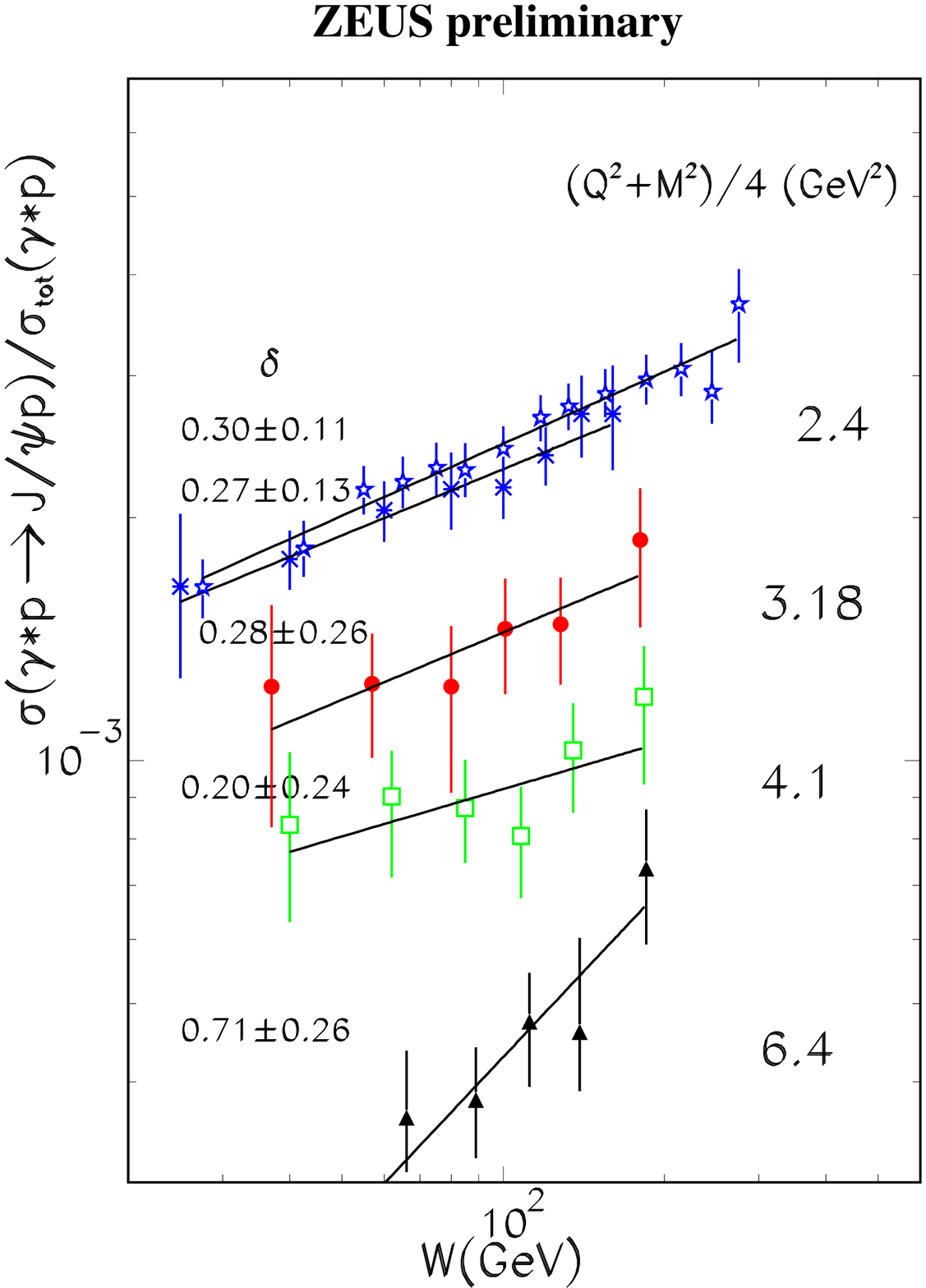}
\caption{The ratio of the $J/\psi$ electroproduction cross section to
the total $\gamma*p$ one, as function of $W$, at different scales. The
lines are a best fit of the form $W^\delta$ to the data.}
\label{fig:ratio-psi}
\end{minipage}
\end{figure}

\section{Deeply Virtual Compton Scattering (DVCS)}

Deeply virtual Compton scattering (DVCS) is a similar process to
electroproduction of VMs, where the final state vector is replaced by
a real photon. The DVCS initial and final states are identical to
those of the QED Compton process. The diagrams of both processes are
shown in Fig. 18.

\vspace{-0.5cm}
\begin{figure}
\begin{center}
\includegraphics[width=.25\textwidth]{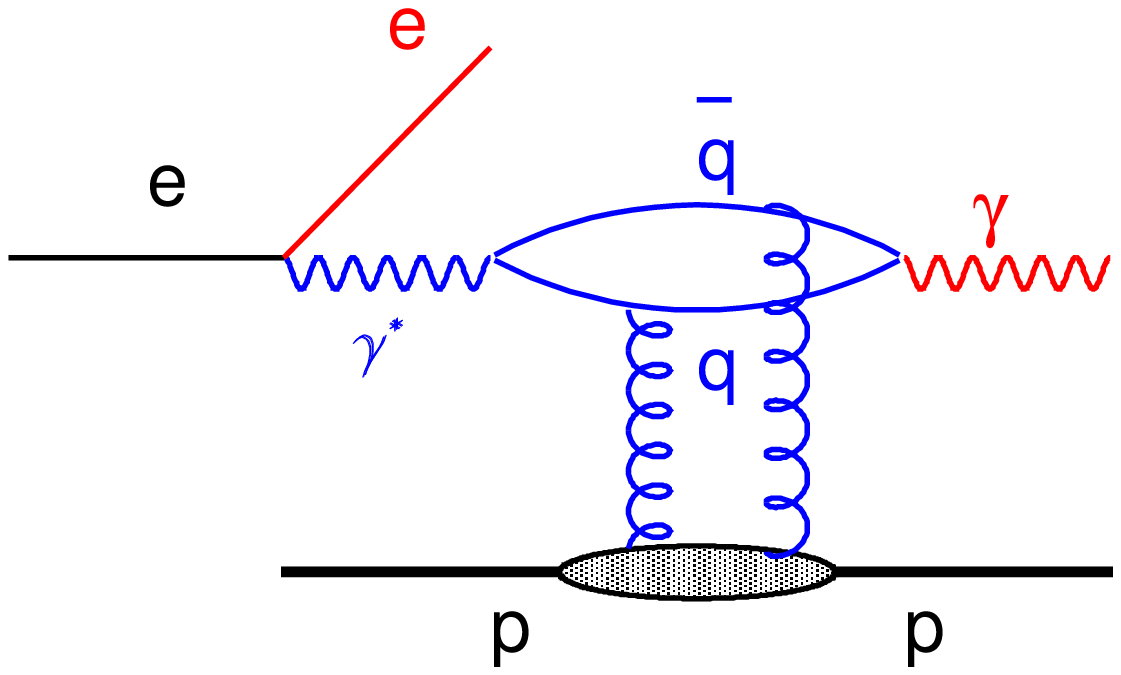}
\includegraphics[width=.25\textwidth]{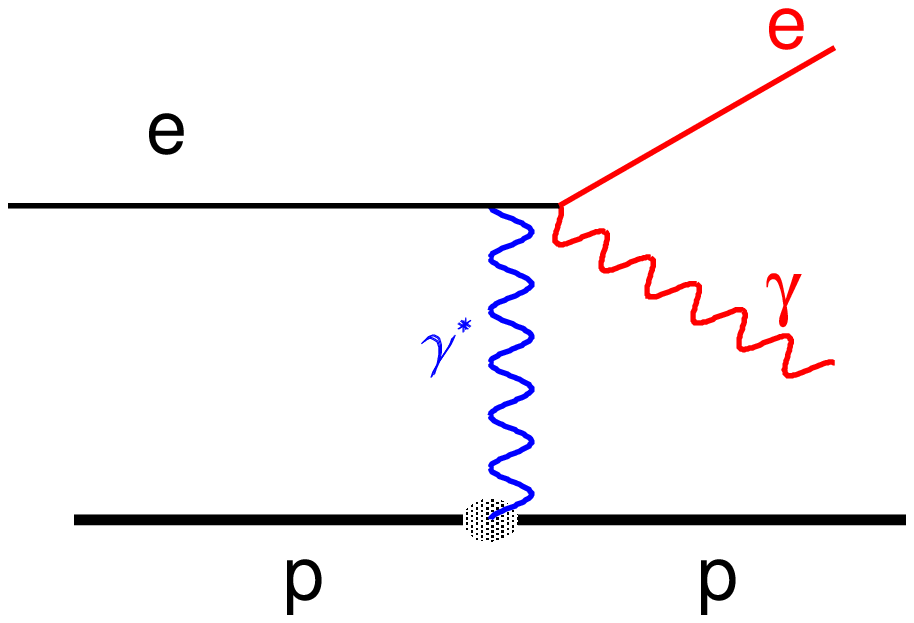}
\includegraphics[width=.25\textwidth]{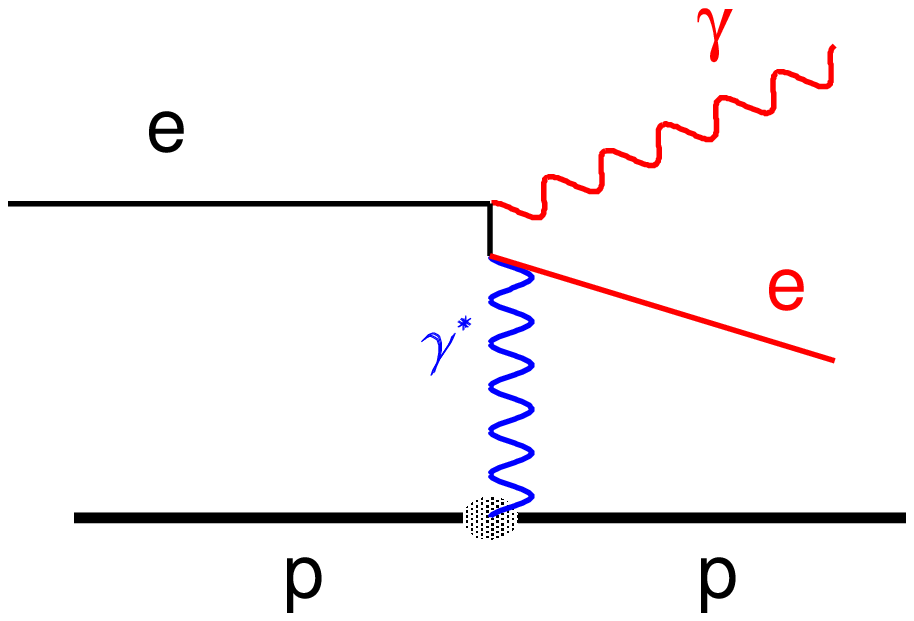}
\end{center}
\caption{Diagrams showing the QCD DVCS process and the QED Compton process.}
\label{fig:dvcs}
\end{figure}

The big interest in DVCS comes from the fact that the QED and QCD
amplitudes interfere and produces an asymmetry which can be measured,
once high statistics data are at hand. This would give information on
the real part of the QCD amplitude. In addition, the DVCS process is a
potential one for obtaining generalized parton
distributions~\cite{gpd}.

\vspace{-0.5cm}
\begin{figure}
\begin{minipage}{5.5cm}
\hspace{-0.9cm}
\includegraphics[width=1.2\textwidth]{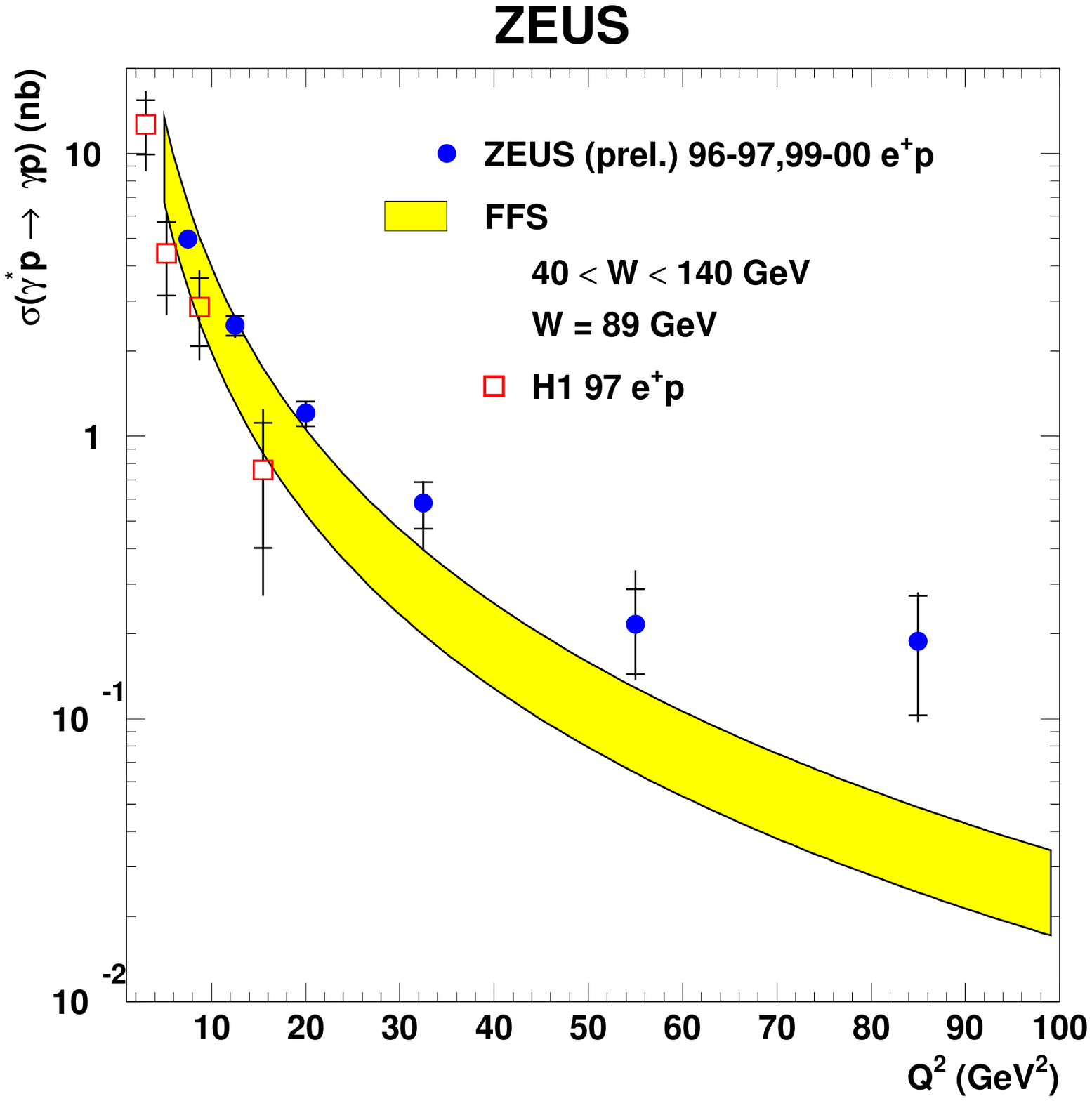}
\caption{The DVCS cross section as function of $Q^2$. The band is a 
theoretical prediction~\cite{ffs}.}
\label{fig:dvcsq2}
\end{minipage}
\hspace{5mm}
\begin{minipage}{5.5cm}
\includegraphics[width=1.2\textwidth]{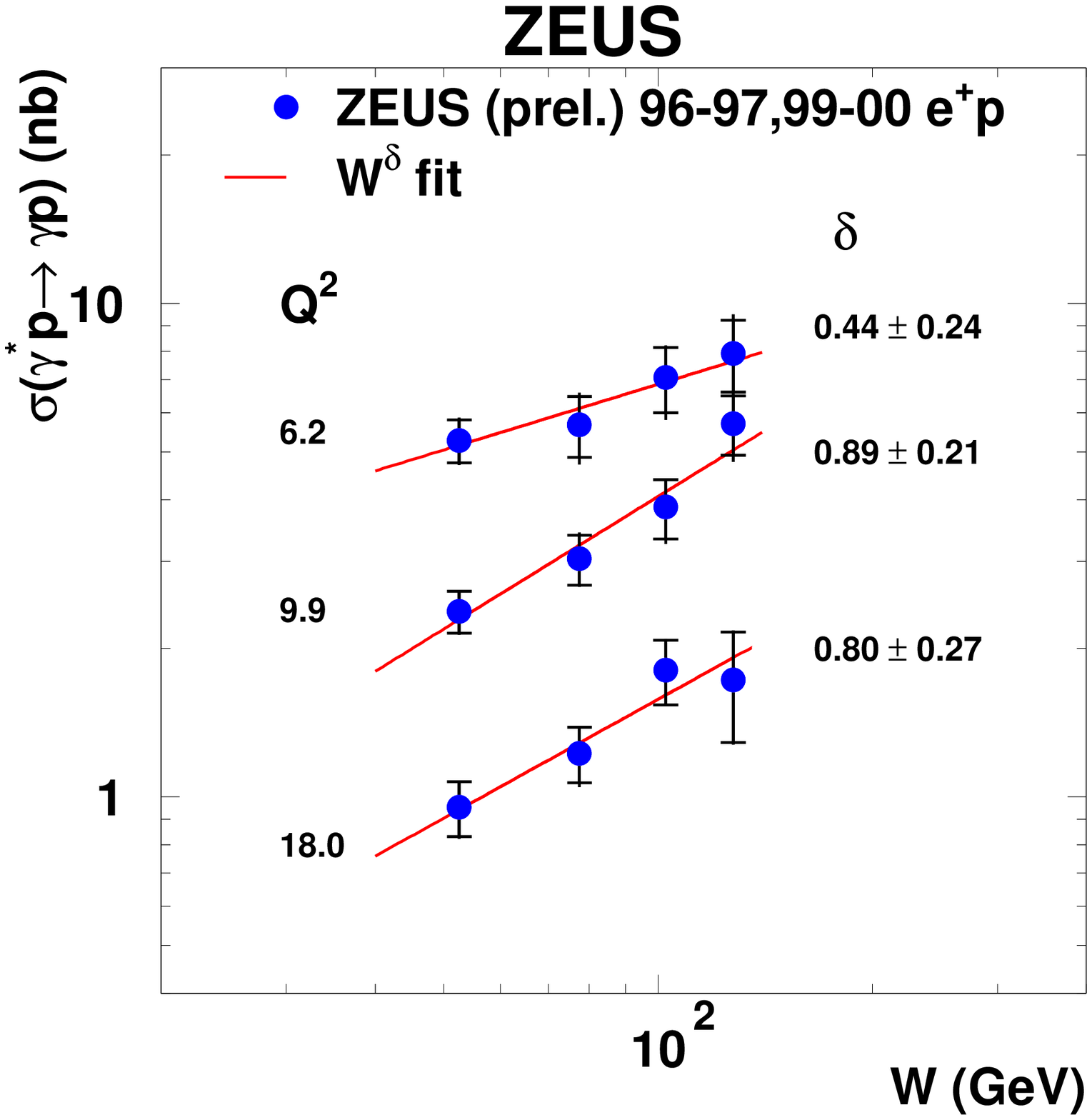}
\caption{The DVCS cross section as a function of $W$ for fixed $Q^2$ values.
  The line is a fit of the form $W^\delta$ to the data.}
\label{fig:dvcsw}
\end{minipage}
\end{figure}

The $Q^2$ dependence of the DVCS cross section is shown in Fig.
19~\cite{zeus-dvcs,h1-dvcs}. It fall-off with $Q^2$ is well described
by the Frankfurt, Freund and Strikman model~\cite{ffs}. The $W$
dependence of the DVCS cross section is shown in Fig.
20~\cite{zeus-dvcs}, for fixed $Q^2$ values. Fitting the data to a
form of $W^\delta$ shows that the cross section rises steeply as $Q^2$
increases. It reaches the same value of $\delta$ as in the hard
process of $J/\psi$ electroproduction. Given the fact that the final
state photon is real, and thus transversely polarized, the DVCS
process is produced by transversely polarized virtual photons,
assuming s-channel helicity conservation. The steep energy dependence
thus indicates that the large configurations of the virtual transverse
photon are suppressed. This is the same conclusion as we obtained
above in the case of the electroproduction of $\rho^0$.

\section*{Acknowledgments}

I would like to thank the organizers for a very pleasant conference.
I would also like to acknowledge the partial support of the Israel
Science Foundation (ISF) and the German Israel Foundation (GIF), which
made this contribution possible.

%

\end{document}

%% file: DEFS.tex
\newcommand {\pom} {I\!\!P}
\newcommand {\pomsub} {{\scriptscriptstyle \pom}}

\newcommand {\xpom} {x_{\pomsub}}
\newcommand {\apom} {\alpha_{\pomsub}}

\newcommand {\aprime} {\alpha^\prime_\pomsub}






%% file: author.bbl
\begin{thebibliography}{8.}
  \addcontentsline{toc}{section}{References}
\bibitem{regge}P. D. Collins, ``An Introduction to Regge Theory and
High-Energy Physics'', Cambridge University Press, 1977.
\bibitem{gribov}V. N. Gribov, JEPT Lett. {\bf 41} (1961) 667.
\bibitem{dl}A. Donnachie and P. V. Landshoff, Phys. Lett. {\bf B296}
(1992) 227.
\bibitem{low-nussinov}F. E. Low, Phys. Rev. {\bf D12} (1975) 163;
S. Nussinov, Phys. Rev. Lett. {\bf 34} (1975) 1286.
\bibitem{ha-ferrara}See e.g. H. Abramowicz, Nucl. Phys. {\bf B}
(Proc. Suppl.) {\bf 99} (2001) 79, and references therein.
\bibitem{collins}J. C. Collins, Phys. Rev. {\bf D57} (1998) 3051;
erratum Phys. Rev. {\bf D61} (1998) 2000.
\bibitem{berera}A. Berera and D. E. Soper, Phys. Rev. {\bf D53} (1996)
6162.
\bibitem{trentadue}L. Trentadue and G. Veneziano, Phys. Lett. {\bf
B323} (1994) 201.
\bibitem{ingelman}G. Ingelman and P. E. Schlein, Phys. Lett. {\bf
B152} (1985) 256.
\bibitem{zeus-f2d3}ZEUS Collab., paper \#823 submitted to ICHEP02,
Amsterdam 2002.
\bibitem{h1-pom}H1 Collab., paper \#980 submitted to ICHEP02,
Amsterdam, 2002.
\bibitem{allm97}H. Abramowicz and A. Levy, DESY 97-251 (1997).
\bibitem{h1-jet}H1 Collab., paper \#987 submitted to ICHEP02,
Amsterdam, 2002.
\bibitem{zeus-dstar}ZEUS Collab., paper \#832 submitted to ICHEP02,
Amsterdam 2002.
\bibitem{fs}L. Frankfurt and M. Strikman, hep-ph/9907221, (1999).
\bibitem{zeus-mx}ZEUS Collab., paper \#821 submitted to ICHEP02,
Amsterdam 2002.
\bibitem{brodsky}S. J. Brodsky et al., Phys. Rev. {\bf D50} (1994) 3134.
\bibitem{zeus-rho}A. Kreisel for the ZEUS Collab., Proceedings of
DIS01, Bologna, 2001.
\bibitem{h1-rho}H1 Collab., paper \# 989 submitted to ICHEP02,
Amsterdam, 2002.
\bibitem{zeus-psi}ZEUS Collab., paper \#813 submitted to ICHEP02,
Amsterdam, 2002.
\bibitem{h1-psi}H1 Collab., C.Adloff et al., Eur. Phys. J {\bf C10}
(1999) 373.
\bibitem{afs}H. Abramowicz, L. Frankfurt and M. Strikman, Surveys High
Energy Phys. {\bf 11} (1997) 51.
\bibitem{align-jet}J. Bjorken, AIP Conference Proceedings No. 6,
Particles and Fields Subseries No. 2, ed. M. Bander, G. Shaw and
D. Wong (AIP, New York, 1972).
\bibitem{zeus-rq2}ZEUS Collab., paper \#818 submitted to ICHEP02,
Amsterdam 2002.
\bibitem{h1-rq2}H1 Collab., paper \#989 submitted to ICHEP02,
Amsterdam 2002.
\bibitem{mrt}A. D. Martin, M. G. Ryskin and T. Teubner,
Phys. Rev. {\bf D55} (1997) 4329; Phys. Rev. {\bf D56} (1997) 3007; 
Phys. Rev. {\bf D62} (2000) 014022.
\bibitem{zeus-ratio}ZEUS Collab., paper \#820 submitted to ICHEP02,
Amsterdam 2002.
\bibitem{gpd}See e.g. A. V. Radyushkin, Phys. Rev. {\bf D58} (1998) 114008.
\bibitem{zeus-dvcs}ZEUS Collab., paper \# 825 submitted to ICHEP02,
  Amsterdam, 2002.
\bibitem{h1-dvcs}H1 Collab., C. Adloff et al., Phys. Lett. {\bf B517}
  (2001) 47.
\bibitem{ffs}L. Frankfurt, A. Freund and M. Strikman, Phys. Rev. {\bf
    D58} (1998) 114001; erratum Phys. Rev. {\bf D59} (1999) 119901.

\end{thebibliography}
